# Distributed Security Constrained Economic Dispatch

M. Hadi Amini[1,2,4], Rupamathi Jaddivada[1], Sakshi Mishra[3], and Orkun Karabasoglu[2,4]

[1] Department of Electrical and Computer Engineering, Carnegie Mellon University, Pittsburgh, PA, USA
[2] SYSU-CMU Joint Institute of Engineering, Guandong, China
[3] Energy Science, Technology and Policy, CIT interdisciplinary, Carnegie Mellon University, Pittsburgh, PA, US
[4] SYSU-CMU Shunde International Joint Research Institute, Guandong, China

Emails: amini@cmu.edu, rjaddiva@andrew.cmu.edu, sakshimi@andrew.cmu.edu, karabasoglu@cmu.edu

*Abstract*— In this paper, we investigate two decomposition methods for their convergence rate which are used to solve security constrained economic dispatch (SCED): 1) Lagrangian Relaxation (LR), and 2) Augmented Lagrangian Relaxation (ALR). First, the centralized SCED problem is posed for a 6-bus test network and then it is decomposed into subproblems using both of the methods. In order to model the tie-line between decomposed areas of the test network, a novel method is proposed. The advantages and drawbacks of each method are discussed in terms of accuracy and information privacy. We show that there is a tradeoff between the information privacy and the convergence rate. It has been found that ALR converges faster compared to LR, due to the large amount of shared data.

## I. INTRODUCTION

An optimization problem is required to be solved for conducting Security Constrained Economic Dispatch (SCED) of the power system. Currently it is solved via approaches in a centralized manner, which is performed by central coordinators. A key feature of future power systems is the utilization of minimal communication in order to ameliorate power systems' performance [1]. In the SCED, the main objective is to find generation dispatch of $N_g$ generators which minimizes the total cost of generation, given a specific demand, denoted by $P_{load}$, taking the physical constraints of a power network into account [2][3]. According to [1], one of the emergent requirements of future power systems is to utilize distributed algorithms by means of data exchange among several entities. Ilić *et al* provide a panorama of distributed power systems in order to model and simulate complex dynamic systems [4]. A comprehensive survey on the existing distributed algorithms to solve the power dispatch problems, i.e., Economic Dispatch and optimal power flow (OPF), is presented in [5]. Kar *et al* also present a novel nodal-based distributed algorithm to tackle the energy management and power dispatch problems [5]. Furthermore, [6] elaborates more on the convergence properties of the distributed method to solve OPF presented in [5]. The proposed method finds a distributed solution for the first order optimality conditions, which thus reduces the optimization problem into solving a coupled system of equations. In this regard, [7] improves the convergence of the method presented in [5] by adding a few additional communication links. Augmented Lagrangian Relaxation (ALR) decomposition is utilized in [8] to solve the security-constrained unit commitment in a decentralized manner. In [9], a system of systems framework is proposed to solve distributed AC optimal power flow for active distribution networks.

The SCED optimization problem can be represented by a constrained optimization problem and can be solved utilizing Lagrange function, i.e. $\min_{P_g} \sum_{i=1}^{N_g} F_i(P_{gi})$, $s.t. \sum_{i=1}^{N_g} P_{gi} = P_{load}$; where $F_i(P_{gi})$ is the cost function of generation unit $i$. Several mathematical approaches such as linear programming have been used to tackle this problem [10]. In addition, we consider the generation limits as well as maximum line flow limits to achieve feasible solution for SCED. In order to account for the maximum line flow limits, line flows have to be determined by conducting power flow study as the first step. In this project, we use DC power flow (DCPF). It is a simplified version of power flow, but still leads to fairly accurate results. There have been several studies, which concentrated on DC power systems and DC microgrids and their effects on the system parameters, such as voltage characteristics [11]. In [12], energy management in DC microgrids and hybrid DC microgrids is utilized as a means to facilitate renewable resources integration. Ref. [13], provides a comprehensive survey on DC microgrids. The advantages of DCPF include, but are not limited to simple implementation, capability to be represented as a linear model, being non-iterative, having reliable and unique solutions, and providing reasonable accuracy for active power flows. Nonetheless, the size of problem in real world power systems is too large which creates a need to divide the problem into smaller sub-problems, which are easier to solve and have lower computational burden. Hug *et al* introduced methods and software for power system distributed management in [14]. We decompose the problem into *r* sub-problems. For sub-problem *i* (area) we have $N_{g,i}$ generators to be dispatched and total number of buses is $N_i$. Figure 1 shows the test network which includes two areas connected via tie-line. In this paper we assume that there are no generators and loads connected at the boundary buses, which are joined by the tie-line. We investigate the convergence rate of both ALR and LR method via this test system.

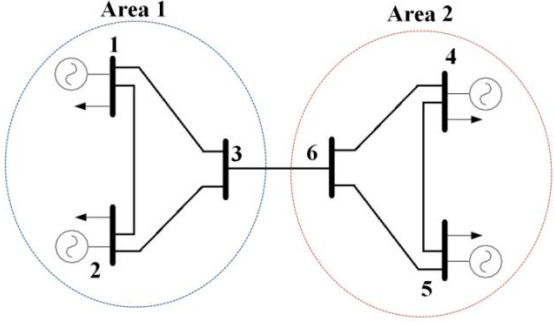

Figure.1. Test network

The rest of this paper is organized as follows: section II provides the general centralized problem formulation of SCED. Section III elaborates each of the decomposition techniques. Afterwards, section IV considers the test network and implements both the decomposition techniques, the results of which are compared and analyzed in Section V. Finally, section VI concludes the paper.

## II. PROBLEM FORMULATION

In this paper, we are interested in the Security Constrained economic dispatch problem, which is a standard minimization problem of the form which is shown is (1).

$$\min_x f(X),$$
$$\text{s.t. } g(X) = 0 \quad (1)$$
$$h(X) \leq 0$$

where $X$ is the vector of control variables and state variables, $f(X)$ is the objective function, $g(X)$ and $h(X)$ are the equality and inequality constraints respectively. In general, the control variables are the active power outputs and reactive power outputs, the state variables are the voltage magnitudes and angles at each of the buses. The objective of SCED is to find out the generator outputs at each of the buses, which would minimize the total cost of generation subject to the physical constraints of the system. In order to obtain these physical constraints, we perform DCPF. Hereafter we use the following notations for the parameters and variables: $\Omega_{sys}$ (set of buses in the system), $\Omega_G$ (set of buses to which a generator is connected), $\Omega_{L_i}$ (set of lines in the system incident on $i^{th}$ bus), $\Omega_D$ (set of buses to which load/ demand is connected), $P_{G_i}$ (output of generator at bus $i$), $P_{D_i}$ (Demand at bus $i$), $C_i(P_{G_i})$ (quadratic cost function of generator at bus $i$), $B_{ij}$ (the element in the susceptance matrix in $i^{th}$ row and $j^{th}$ column), $\theta_i$ (voltage angle at the $i^{th}$ bus), and $F_{ij}$ (flow in the line joining $i^{th}$ bus and $j^{th}$ bus).

*Objective function*: The objective function for the SCED problem is the cost of generation of the entire system. The cost of the generator is modeled as a quadratic cost function given by $C_i(P_{G_i}) = a_i + b_i P_{g_i} + c_i P_{g_i}^2$; where, $a_i$, $b_i$ and $c_i$ are the cost coefficients of the $i^{th}$ generating unit. Hence, the objective function is represented by $f(X) = \sum_{i \in \Omega_G} C_i(P_{g_i})$ in $/h$.

*Equality constraints*: In the case of contingency, there might not be enough time to carry out the entire power flow. In such a situation, the DCPF method gives a solution, which could be utilized to stabilize the system in order to ensure security of the system. Consequently, for each bus we can write $g(X) = P_{g_i} - P_{L_i} - \sum_{j \in \Omega_{L_i}} F_{ij}$; where $i$ represents the bus number. These set of equations represent power flow balance at each bus, i.e. $\forall i \in \Omega_{sys}$. Furthermore, in the case of DCPF, the flow in the lines, denoted by $F_{ij}$, is given by $F_{ij} = B_{ij} \times (\theta_i - \theta_j)$.

*Inequality constraints*: The SCED problem formulation involves limits on the generation and line flows which can all be considered as inequality constraints. These inequalities can be formulated as (2).

$$\begin{aligned} P_{G_i}^{min} \leq P_{G_i} \leq P_{G_i}^{max} & \quad \forall i \in \Omega_G \\ |F_{ij}| \leq F_{ij}^{max} & \quad \forall i,j \in \Omega_{sys} \end{aligned} \quad (2)$$

*Decomposition into subproblems*: The optimal solution of the entire power system in a centralized manner is not only computationally expensive but very time-consuming. It also requires different regions to exchange all the data required to solve the optimization problem. Owing to this and many other factors, there is a necessity to divide the entire system into sub systems such that the SCED problem can be solved for each of the sub systems separately in a distributed way. The constraints linking different sub-problems are then taken into account and the master problem is solved. There are many techniques discussed in the literature to utilize decomposition techniques to facilitate such optimization [17][18]. In [19], the discrete ED problem is formulated as a knapsack problem. Xu et al utilized a distributed dynamic programming to solve ED problem in a distributed manner.

In order to apply any of the decomposition technique, there first arises the need to divide the system into sub systems and formulate SCED problem for each of the sub systems and define the complicating constraints (constraints which require information from multiple subsystems) for the problem as a whole. Perhaps, the easiest way to decompose the huge power system network is to divide it into regions which are interconnected by tie lines. This is not only because of weak coupling but also caused by the minimum communication requirement between these subsystems. Next section introduces the elaborate formulation of the two utilized decomposition techniques.

## III. DECOMPOSITION TECHNIQUES

### A. Lagrangian Relaxation Decomposition Method

Lagrangian Relaxation (LR) decomposition technique is used for solving a non-linear problem with decomposable structure, which involves complicating constraints. Complicating constraints are constraints that if relaxed, the resulting problem decomposes into simpler problems [15]. For SCED, the problem formulation is described in the previous section. The *non-complicating* equality constraints written as $g(X) = P_{g_i} - P_{L_i} - \sum_{j \in \Omega_{L_i}} F_{ij}; \forall i \in \Omega_{sys}, i \notin \Omega_{tie}$.

We formulate non-complicating inequality constraints as (3).

$$\begin{aligned} h_1(X) = P_{G_i}^{min} \leq P_{G_i} \leq P_{G_i}^{max}; \forall i \in \Omega_G, i \notin \Omega_{tie} \\ h_2(X) = |F_{ij}| \leq F_{ij}^{max}; \forall i,j \in \Omega_{sys,k}, i,j \notin \Omega_{tie} \end{aligned} \quad (3)$$

In addition, the complicating equality and complicating inequality constraints respectively are given by (4).

$$g_{int}(X) = P_{g_i} - P_{L_i} - \sum_{j \in \Omega_{L_i}} F_{ij}, \forall i \in \Omega_{tie}$$
$$h_{int}(X) = |F_{ij}| \leq F_{ij}^{max}; \forall i,j \in \Omega_{tie} \quad (4)$$

We define Lagrangian function as shown in (5).
$$\mathcal{L}(P_{G_i}, \lambda, \mu) = f(X) + \lambda^T g_{int}(X) + \mu^T h_{int}(X); \forall i \in \Omega_G \quad (5)$$
where $\lambda$ and $\mu$ are the Lagrange Multiplier vectors for equality and inequality constraints respectively and $g_{int}(X)$ and $h_{int}(X)$ are set of complicating equality and complicating inequality constraints respectively.

Under regularity and convexity assumptions, the dual function (DF) is defined as shown in (6).
$$\phi(\lambda, \mu) = \min_x \mathcal{L}(P_{G_i}, \lambda, \mu); \forall i \in \Omega_G$$
$$\text{s.t. } g(X) = 0; h_1(X) \leq 0; h_2(X) \leq 0 \quad (6)$$

The dual function is concave and in general non-differentiable. This fundamental fact is exploited in the Lagrangian Relaxation Decomposition algorithm. The dual problem is then defined as $\text{Max}_{\lambda,\mu} \phi(\lambda, \mu)$ s.t. $\mu \geq 0$.

We formulate the $k^{th}$ subproblems as shown in (7).
$$\min_{x_k} f_k(X_k) + \lambda^T g_{int,1}(X_k)$$
$$\text{s.t. } g_k(X_k) = 0 \quad (7)$$
$$h_k(X_1) \leq 0$$

*Formulation of the decomposed system:* The primal SCED problem is decomposed into $r$ sub-problems based on the geographical regions and utility ownership of the physical system. Let $\Omega_{tie,r}$ denote the set of buses in the $r^{th}$ subsystem that has a tie line incident on it. The SCED problem can be formulated for $k^{th}$ subproblem as shown in (8).

$$\min_{P_{G_i}} \sum_{i \in \Omega_{G,k}} \mathcal{L}(P_{G_i}, \lambda, \mu)$$
$$\text{s.t. } P_{G_i}^{min} \leq P_{G_i} \leq P_{G_i}^{max} \quad \forall i \in \Omega_{G,k}$$
$$P_{g_i} - P_{L_i} - \sum_{j \in \Omega_{L_i}} F_{ij} = 0; \forall i \in \Omega_{sys,k}, i \notin \Omega_{tie,k} \forall k = 1,2,\dots r \quad (8)$$
$$|F_{ij}| \leq F_{ij}^{max} \quad \forall i,j \in \Omega_{sys,k}, i,j \notin \Omega_{tie,k} \forall k = 1,2,\dots r$$

In the decomposed formulation, the objective function comprises the cost functions for all the generators pertaining to $k^{th}$ region. The physical constraints comprising of parameters belonging to more than one sub-problem are the coupling constraints for the system. If there is a tie line between $p^{th}$ region and $q^{th}$ region, the coupling constraints can be taken into account as shown in (9).

$$\begin{cases} P_{g_i} - P_{L_i} - \sum_{j \in \Omega_{L_i}} F_{ij} = 0 & \forall i \in \{\Omega_{tie,q}, \Omega_{tie,p}\} \\ |F_{ij}| \leq F_{ij}^{max} & \forall i \in \Omega_{tie,p}, j \in \Omega_{tie,q} \end{cases} \quad (9)$$

*Four steps of the algorithm [15][16]:* 1) The Largrange Multiplier vector is initialized; 2) The subproblems are solved; 3) The values of Lagrange Multipliers are updated; and 4) If stopping condition is fulfilled, the procedure terminates, otherwise the subproblems are solved again by going to step 2.

In this paper, we use subgradient method for updating the Lagrange Multipliers [15]. Assume that $s^{(v)}$ denotes the mismatch vector of coupling constraints at iteration $v$, i.e., $s^{(v)} = \begin{bmatrix} g_{int}(x^{(v)}) \\ h_{int}(x^{(v)}) \end{bmatrix}$. We update the multipliers using (10).

$$\begin{bmatrix} \lambda^{(v+1)} \\ \mu^{(v+1)} \end{bmatrix} = \begin{bmatrix} \lambda^{(v)} \\ \mu^{(v)} \end{bmatrix} + k^{(v)} \frac{s^{(v)}}{||s^{(v)}||} \quad (10)$$

where $k^{(v)} = \frac{1}{a+bv}$. If $\mu^{(v+1)} < 0$ then $\mu^{(v+1)} = 0$.

*B. Augmented Lagrangian Relaxation Method*

ALR decomposition technique is used for solving non-linear problems with decomposable structure. This approach has been widely used in power system optimization problems, for instance, Beltran *et al* utilized ALR decomposition to solve unit commitment problem. According to [17], one of the drawbacks of ALR method is the introduction of a non-separable quadratic term, which requires few additional methods such as auxiliary problem principle to deal with this non-separability of the corresponding Lagrangian function. An interesting study on the application of ALR in multi-area optimal power flow is performed by Conejo and his colleagues in 2003, which highlighted the advantages of this method. According to [18], utilizing ALR to solve OPF problem will ameliorate computation efficiency. For SCED, the problem formulation is described in the previous section. With the same equations, we only classified equality constraints into $g(X) = 0$ as non-coupling and $g_{int}(X) = 0$ as coupling constraints in addition. The Lagrangian function is defined as (11).

$$\mathcal{L}_{AL}(X, \lambda, \mu) = f(X) + \lambda^T g_{int}(X) + \frac{1}{2}\gamma ||g_{int}(X_1, X_2)||^2 \quad (11)$$

Considering the fact that we have two subproblems for our test system, $g(X) = 0$ can be shown as $g_1(X_1) = 0$ and $g_2(X_2) = 0$ which correspond to subproblems one and two, respectively. We can apply this to the noncoupling inequality constraints as well to determine the inequality constraint related to each sub-problem. In this paper, we keep coupling equality constraints as it is in the Lagrange function and then use directly fixing variables approach, which is called the *alternating direction method*. The determined subproblems are shown in (12), by writing the objective function as combination of two separable objective functions of each of the subproblems. i.e., $f(X) = f_1(X) + f_2(X)$. Here is the sequence of solving subproblems: 1) minimize the first subproblem by assigning some initial value to the variables of other subproblems, 2) solve the second subproblem by substituting the variables of the first subproblem with the ones determined in the previous step, 3) update the Lagrange multipliers accordingly.

Step1) $\min_{x_1} f_1(X_1) + \lambda^T g_{int,1}(X_1) + \frac{1}{2}\gamma ||g_{int}(X_1, X_2^k)||^2$
$\text{s.t. } g_1(X_1) = 0, h_1(X_1) \leq 0$

Step2) $\min_{x_2} f_2(X_2) + \lambda^T g_{int,2}(X_2) + \frac{1}{2}\gamma ||g_{int}(X_1^k, X_2)||^2 \quad (12)$
$\text{s.t. } g_2(X_2) = 0, h_2(X_2) \leq 0$

Step3) $\lambda^{k+1} = \lambda^k + \alpha \cdot g_{int}(X_1^{k+1}, X_2^{k+1})$

Modeling the coupling inequality constraints with ALR is not straightforward. Hence, we changed the topology of our problem formulation to convert the shared inequality constraint related to the tie-line which connects the two areas, to two non-coupling inequality constraints in each sub-problem separately, i.e., we add the constraints shown in (13) to subproblem (1) and subproblem (2) respectively.

1) $|\sum_{i\in\Omega_{G,1}} P_{g_i} - \sum_{i\in\Omega_{L,1}} P_{Di}| \leq F_{tie_{max}}$
2) $|\sum_{i\in\Omega_{G,2}} P_{g_i} - \sum_{i\in\Omega_{L,2}} P_{Di}| \leq F_{tie_{max}}$ (13)

These additional constraints ensure that the tie-line flow does not exceed the limit. The flowchart of implementing ALR for a two-area network is shown in Fig. 2.

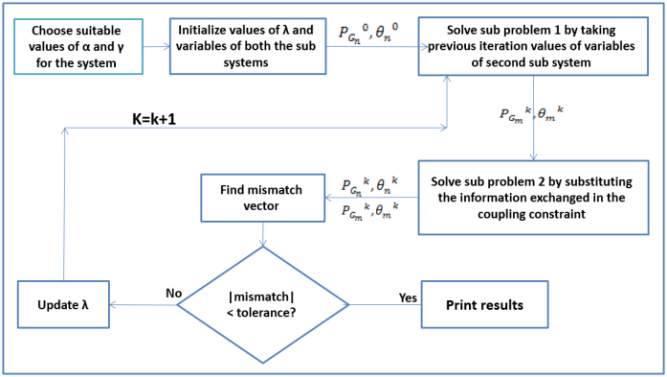

Figure 2. Flowchart of ALR Decomposition

## IV. CASE STUDY

### A. LR Decomposition Results

Total load of the test system is 1.2 MW distributed equally at buses 1, 2, 4, and 6. The results of solving the security constrained economic dispatch for two area, 6-bus test system considering a=3, b=0.2 , for using in the subgradient method of updating Lagrange multiplier as discussed in Section III, are:
$P_{g1} = 0.1, P_{g2} = 0.2841, P_{g4} = 0.4121, P_{g5} = 0.3797$

It takes 10.558829 seconds to solve this optimization problem in 281 iterations.

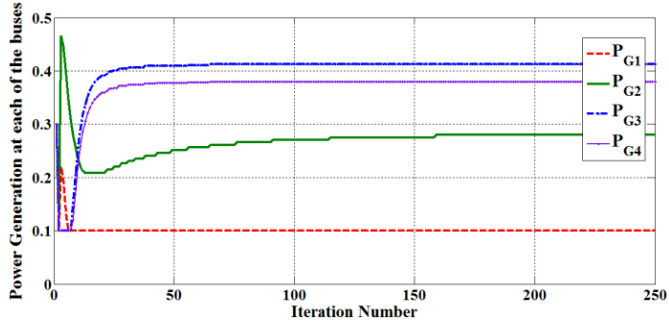

Figure 3. Solutions obtained through LR

***Sensitivity analysis to tune the optimization parameters:*** The effect of the tuning parameters on the number of iterations to converge, total run-time and error value are represented in the table I. According to this table, in cases 1-4 we fix the value of ***b*** and did a sensitivity analysis on the error values with respect to changes of ***a***. We see that error does not vary linearly with . By decreasing ***a*** from 4 to 1, the error initially reduced and then increased. As a result, we select ***a=2*** as the best value among the tested values.

Table I. Sensitivity analysis of the parameters

| Case | 1 | 2 | 3 | 4 | 5 | 6 | 7 | 8 | 9 |
|---|---|---|---|---|---|---|---|---|---|
| a | 4 | 3 | 2 | 1 | 0.5 | 3 | 3 | 3 | 2 |
| b | 0.2 | 0.2 | 0.2 | 0.2 | 0.2 | 0.3 | 0.1 | 0.05 | 0.05 |
| Iteration | 246 | 281 | 225 | 130 | 672 | 867 | 79 | 63 | 48 |
| Time (sec.) | 8.4 | 10.5 | 8.3 | 4.4 | 23.6 | 31.0 | 2.8 | 2.3 | 2.0 |
| Error | 2.2% | 2% | 1.9% | 2.2% | 2.0% | 1.9% | 2.2% | 1.9% | 1.8% |

Afterwards, in cases 5-8, we fix the value of ***a*** and did a sensitivity analysis on the error values with respect to changes in ***b***. Following a similar approach which is implemented for cases 1-4, we see that also doesn't have a linear relation with the error. By decreasing ***b*** from 4 to 1, the error first reduced, then increased and eventually it reduced again. Additionally, we should also consider the number of iterations. Between our two candidates for ***b*** value, which are 0.05 and 0.3, we select the case which has less number of iterations. As the following figure shows, the number of iterations and run-time are directly related, i.e. by increasing the number of iterations, the run-time will increase proportionally. In this case, the number of iterations for ***b=0.05*** is considerably less than that for ***b=0.3***. Hence, we select this value as the best value for ***b*** among the tested values.

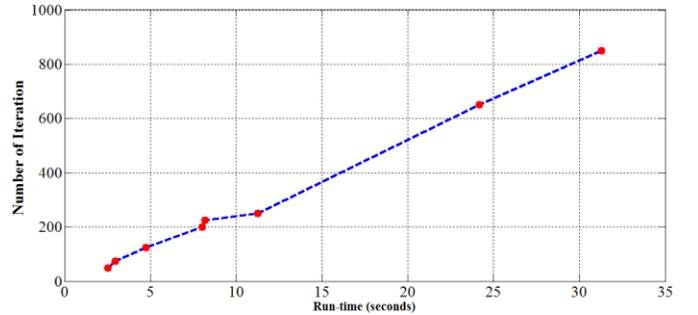

Figure 4. Run time vs number of iterations required for convergence

To validate this claim, we can eventually adjust the tuning parameter such that we achieve more accurate results in less time. We solve the optimization with the values determined by case 1-4 and cases 5-8 for ***a*** and ***b*** respectively. The results represents that both error value and the number of iteration in the last case is less than all previous cases (cases 1-8).

### B. Augmented Lagrangian Relaxation Method

Taking the initial values of tuning parameters as $\alpha = 0.1$, $\gamma = 0.15$ and the initial Lagrange multiplier vector to account for each of the coupling constraints as $\lambda = \begin{bmatrix} 3.066 \\ 3.066 \end{bmatrix}$, the following results were obtained: $P_{g1} = 0.1, P_{g2} = 0.2856, P_{g4} = 0.4161, P_{g5} = 0.3838$

***Sensitivity analysis to tune the optimization parameters:***

The selection of tuning parameters is an integral part of any optimization problem. In the following lines, the sensitivity analyses of the parameters are done via parameter tuning for α and γ.

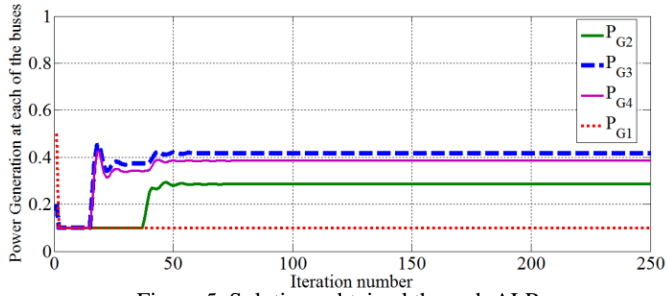

Figure 5. Solutions obtained through ALR

- γ : γ is made to vary from 0.05 to 0.75 in steps of 0.05. The initial values of γ results in oscillations and it doesn't converge to a feasible solution. As it is further increased, the values converge and there is more or less no change in the accuracy. Further, the number of iterations required for convergence decrease and then increase. The optimum value of γ thus obtained is 0.25, resulting in faster convergence and more accuracy. The aforementioned discussion will be clear through fig. 6.

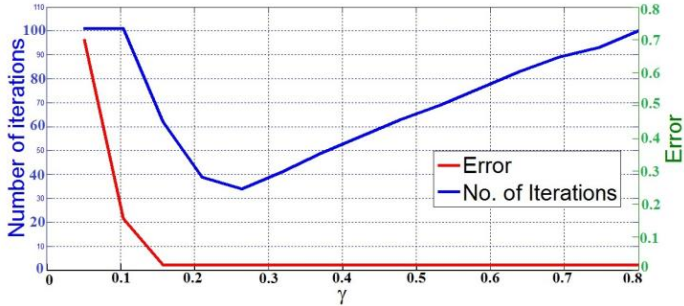

Figure 6. Sensitivity analysis of the parameter γ

- α: This is the parameter used for the Lagrange multiplier update. Larger values beyond 0.2 results in more oscillations, hence resulting in larger number of iterations for convergence.

- λ: These are made to vary from 2 to 20 at each of the buses. The number of iterations required for convergence reduces greatly from 102 to 36 and then slowly increase to 70. The least number of iterations are obtained for λ chosen between the values 5 and 6 at each of the buses. Also, it is discerned from the decentralized solution that the system's LMPs are 5.74 and 5.11 respectively. Hence initialization of λ to the values between 5 and 6 converges faster.

## V. DISCUSSION AND ANALYSIS

### A. Comparison of execution time for a given number of iterations

Both the methods are run for a fixed number of iterations and the time elapsed has been noted. We observe that the time taken by LR increases linearly with the number of iterations. On the other hand, the time taken by ALR decreases and then again increases as the number of iterations for which it is allowed to run is varied. In an attempt to compare these two techniques, both the plots are superimposed and it is observed that the time taken by LR is always less than the time taken by ALR. This is because LR takes less time to carry out each iteration.

### B. Comparison of the error for a given number of iterations

For the sake of comparison, error is defined as the difference between the total generation and the total load. It has been observed that as the number of iterations for which SCED is allowed to run increases, the error in both the decomposition techniques decreases.

The convergence criteria is now changed to have the absolute value of the mismatch of the power generation and consumption at each time period smaller than 0.01. Both the methods converge with substantially different number of iterations and time. The time taken by LR to converge was 78.99 seconds with 225 iterations and an error of 0.0134. While, ALR converged within 18.7721 seconds, 51 iterations and an error of 0.0033.

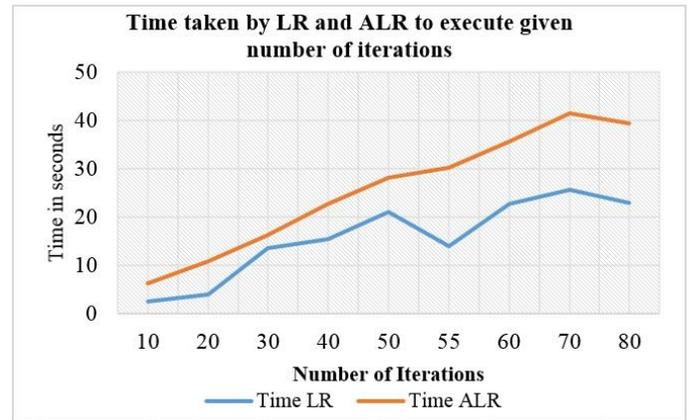

Figure 7. Comparison of LR and ALR in terms of convergence time for given number of iterations

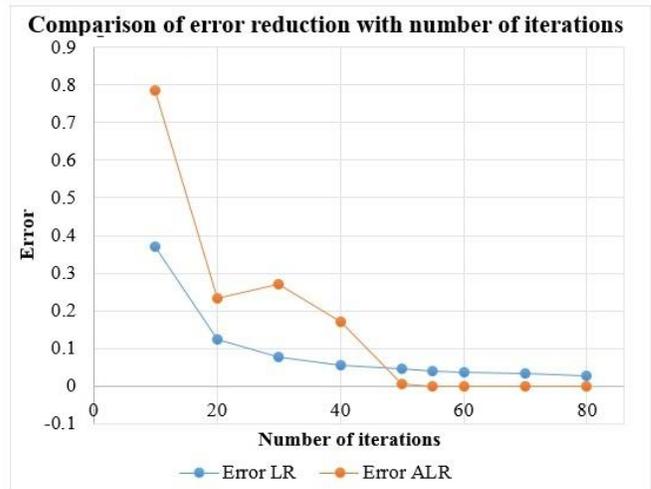

Figure 8. Comparison of LR and ALR in terms of error for specific number of iterations

## VI. CONCLUSION

We utilized two decomposition methods: Lagrangian Relaxation (LR) and Augmented Lagrangian Relaxation (ALR) to solve the Security Constrained Economic Dispatch problem. In order to compare the two methods, we applied both on the same test system which consists of two areas connected via a tie-line. The time taken for LR to converge increases linearly with the number of iterations it is allowed to run for. Moreover, convergence time taken by LR is always less than that of ALR. Additionally, we investigated the effect of parameter tuning in each of the evaluated decomposition techniques on the convergence rate. Through the analysis, we have shown that appropriate tuning of optimization parameter can effectively ameliorate the convergence rate for LR and ALR decomposition methods. As a corollary, we found out that the smaller values of $\gamma$ in ALR can result in oscillations.

From the information privacy perspective, we found that the decomposition techniques reduce the amount of shared information between subproblems (areas). The LR method requires the sharing of the data related to tie-line flow only, i.e., voltage magnitude and angles at the boundary buses. However, a larger amount of data is required to be shared in the case of ALR decomposition technique, i.e., in addition to the information shared in the case of LR method, we also need to share the variables related to the flow of the lines that are incident on the boundary buses in each of the areas. It has been noticed that ALR converges faster compared to LR, which is cause by the large amount of shared data. Consequently, we conclude that there is a tradeoff between the information privacy and the convergence rate, i.e., sharing more data leads to faster convergence. In order to choose appropriate decomposition technique for a specific problem, three factors play a pivotal role: 1) acceptable level of information privacy, 2) convergence rate, and 3) numerical computation cost.

## ACKNOWLEDGMENT

We would like to acknowledge the valuable inputs, fruitful comments, and discussions of Prof. Gabriela Hug to this article.